\documentclass[aps,prd,reprint,amsmath,amssymb,showpacs,superscriptaddress,nofootinbib,twocolumn]{revtex4-2}
\usepackage{multirow}
\usepackage{amssymb}
\usepackage{graphicx}
\usepackage{graphics}
\usepackage{epsfig}
\usepackage{bigstrut}
\usepackage{bm}
\usepackage{xspace}
\usepackage{amsmath}
\usepackage[mathlines]{lineno}
\usepackage{color}
\usepackage[bookmarksnumbered, pdfstartview=FitH,colorlinks,citecolor=blue,linkcolor=blue,]{hyperref}
\usepackage[perpage,symbol]{footmisc}
\usepackage{subfigure}
\usepackage{indentfirst}
\usepackage{fancyvrb}
\usepackage{float}
\usepackage{textpos}
\usepackage[english]{babel}
\renewcommand{\thefootnote}{\fnsymbol{footnote}}

\begin{document}

\title{\bf Novel approach to investigate $\eta$ decays via $\eta^\prime\rightarrow\pi\pi\eta$}

\author{Xiaolin Kang}
\email[]{kangxiaolin@cug.edu.cn}
\affiliation{China University of Geosciences, Wuhan 430074, People's Republic of China}

\author{Yuyao Ji}
\email[]{yuyao.ji@ihep.ac.cn}
\affiliation{Shandong University, Jinan 250100, People's Republic of China}

\author{Xiaoqing Yuan}
\email[]{yuanxq@ihep.ac.cn}
\affiliation{Institute of High Energy Physics, Beijing 100049, People's Republic of China}

\author{Benhou Xiang}
\email[]{xiangbh@ihep.ac.cn}
\affiliation{Institute of High Energy Physics, Beijing 100049, People's Republic of China}
\affiliation{University of Chinese Academy of Sciences, Beijing 100049, People's Republic of China}

\author{Xiaorong Zhou}
\affiliation{University of Science and Technology of China, Hefei 230026, People's Republic of China}
\affiliation{State Key Laboratory of Particle Detection and Electronics, Hefei 230026, People's Republic of China}

\author{Haiping Peng}
\affiliation{University of Science and Technology of China, Hefei 230026, People's Republic of China}
\affiliation{State Key Laboratory of Particle Detection and Electronics, Hefei 230026, People's Republic of China}

\author{Xingtao Huang}
\affiliation{Shandong University, Jinan 250100, People's Republic of China}

\author{Shuangshi Fang}
\affiliation{Institute of High Energy Physics, Beijing 100049, People's Republic of China}
\affiliation{University of Chinese Academy of Sciences, Beijing 100049, People's Republic of China}

\begin{abstract}

To avoid the impact from the background events directly from $e^+e^-$
annihilations or $J/\psi$ decays, we propose a novel approach to investigate
$\eta$ decays, in particular for its rare or forbidden decays, by using
$\eta^\prime\rightarrow\pi\pi\eta$ produced in $J/\psi$ decays at the $\tau-$charm factories. 
Based on the MC studies of a few typical decays, $\eta\rightarrow \pi\pi$,
$\gamma l^+l^- (l= e, \mu)$, $l^+l^-$, as well as $l^+l^-\pi^0$, the sensitivities
could be obviously improved by taking advantage of the extra constraint of $\eta^\prime$.  
Using one trillion $J/\psi$ events accumulated at the Super $\tau$-Charm facility,
the precision on the investigation of $\eta$ decays could be improved significantly
and the observation of the rare decay $\eta\rightarrow e^+e^-$ is even accessable. 
\end{abstract}

\maketitle

\section{Introduction}

Since its strong, electromagnetic, and weak decays are forbidden in the first order,
$\eta$ meson plays an important role as a test of low-energy Quantum Chromodynamics
(QCD) calculations in the framework of chiral perturbation theory (ChPT). 
In addition, $\eta$ is an eigenstate of the charge conjugation ($C$) and parity
($P$) operators, and thus it provides an important experimental tool for investigations
of the degree of conservation of these symmetries in strong and electromagnetic interactions. 
In addition to the promising numbers of $\eta$  directly produced from hadron-production
or photo-production processes, huge samples of the $\eta$ can be collected in the
radiative decays of the vector meson from the $e^+e^-$ annihilations 
($\phi\rightarrow\gamma\eta $ at KLOE-2~\cite{KLOEExp} and $J/\psi\rightarrow\gamma\eta$
at BESIII~\cite{BESIIIExp}).  In recent years, with the world's largest $J/\psi$ samples
collected with the BESIII detector, a series of interesting results on $\eta$ decays was
achieved with the decays of $J/\psi\rightarrow\gamma\eta$ (see the
reviews~\cite{Fang:2017qgz, Fang:2021wes, Fang:2021hyq, Gan:2020aco} for details).

However, it was found that the large background contributions from $J/\psi$ decays
makes it hard to improve the sensitivity for the investigation on the $\eta$ rare
or forbidden decays.  Take $\eta\to\pi^0\pi^0$ as an example, the dominant background
events come from $J/\psi\to\gamma\pi^0\pi^0$ due to the direct pions production.
In particular the production of the intermediate state $f_0(600)$ makes the background
events unreduceable~\cite{BESIII:2011ggk}. To avoid the background impacts directly
from $J/\psi$ decays, we introduce a novel approach to investigate the $\eta$ decays
via $\eta^\prime\to\pi\pi\eta$ process. According to the Particle Data Group (PDG)~\cite{pdg2022},
the producted branching fraction of  $J/\psi\to\gamma\eta^\prime$, $\eta^\prime\to\pi^+\pi^-\eta$
is $(2.23\pm0.04)\times 10^{-3}$, which is about two times larger than that of
$J/\psi\to\gamma\eta$. After taking into account the tracking efficiency of two
charged pions, the selected $\eta$ samples from this approach is larger than,
at least compatible with, the directly obtained sample from $J/\psi\rightarrow\gamma\eta$.
On the other hand, since the $\eta^\prime$ is quite narrow,  one more constraint on the
$\eta^\prime$ peak makes it easier to suppress the background events directly from $J/\psi$ decays. 

Most recently, a project of the Super $\tau$-Charm facility (STCF)~\cite{stcf}
was proposed for exploring the $\tau-$charm physics and searching for the physics
beyond the Standard Model (SM). The STCF is an electron-positron collider, operating
at energies from 2 to 7 GeV, together with a state-of-the-art particle detector.
The designed luminosity, $0.5\times 10^{35}$ cm$^{-2}$s$^{-1}$ or higher is about
100 times larger than that of the BEPCII~\cite{BEPCII}, which enables to collect
unprecedented high statistics data samples in one year. As advocated by the BEPCII/BESIII,
not only will this facility play a leading role in the investigation of $\tau$-charm physics,
but they will offer an unprecedented opportunity to explore the light meson decays
benifitted from the high production rates of light mesons in the charmonium decays.

According to the latest conceptual design report~\cite{stcf}, 3.4 trillion
$J/\psi$ events can be produced in one year. To have a conservation estimation 
on the investigation of $\eta$ decays, the sensitivities are estimated based on
1 trillion $J/\psi$ events, which corresponds to 5.2 billion $J/\psi\rightarrow\gamma\eta'$
decays. Therefore, a simulated sample of 5.2 billion $J/\psi\rightarrow\gamma\eta'$
with $\eta'$ inclusive decays are simulated based on the basic STCF fast simulation
package~\cite{Shi:2020nrf}. All the branching fractions of $\eta'$ decays are taken
from the PDG~\cite{pdg2022}. This sample will be denoted as Pseudo-data throughout
the text and used to estimate the potential background contributions. Then exclusive 
MC studies of a few typical decays of $\eta$ meson are performed in this article to
elucidate the feasibility for investigating $\eta$ decays with $\eta^\prime\rightarrow\pi^+\pi^-\eta$. 
It is worth mentioning that the detector geometry and performance and
the reconstruction software are still under further optimization, such as the spatial
resolution for tracks and clusters, the energy resolution for clusters, the efficiency
for tracking and particle identification.

\section{$\eta\rightarrow\pi\pi$}

The $P$ and $CP$ violating decay $\eta\to\pi\pi$ are usually regarded as the golden
channels to search for the unconventional source of $CP$ violation~\cite{Jarlskog:1995ww}.
The SM and its extended sector predicted the branching fraction of $\eta\to\pi\pi$ 
at the level of $\sim10^{-15}$~\cite{Jarlskog:2002zz}. While the experimental upper
limits are highly limited due to the unreducable background production at both
hadronic collisions and $e^+e^-$ annihilations.
That is why a possible new test in the decay into four pions is performed by 
many experiments even through the detection efficiency is lower than that of 
$\eta\to \pi^0\pi^0$. The present upper limit for branching faction of 
$\eta\rightarrow\pi^0\pi^0$, $3.5\times 10^{-4}$~\cite{Blik:2007ne}, is two 
order magnitudes larger than that of $\eta\rightarrow 4\pi^0$. 
While the upper limit for branching faction of $\eta\rightarrow\pi^+\pi^-$
is $4.4\times 10^{-6}$~\cite{KLOE-2:2020ydi} from KLOE-2 experiment.

\begin{figure}[!htbp]
\centering
\includegraphics[angle=-90,width=0.4\textwidth]{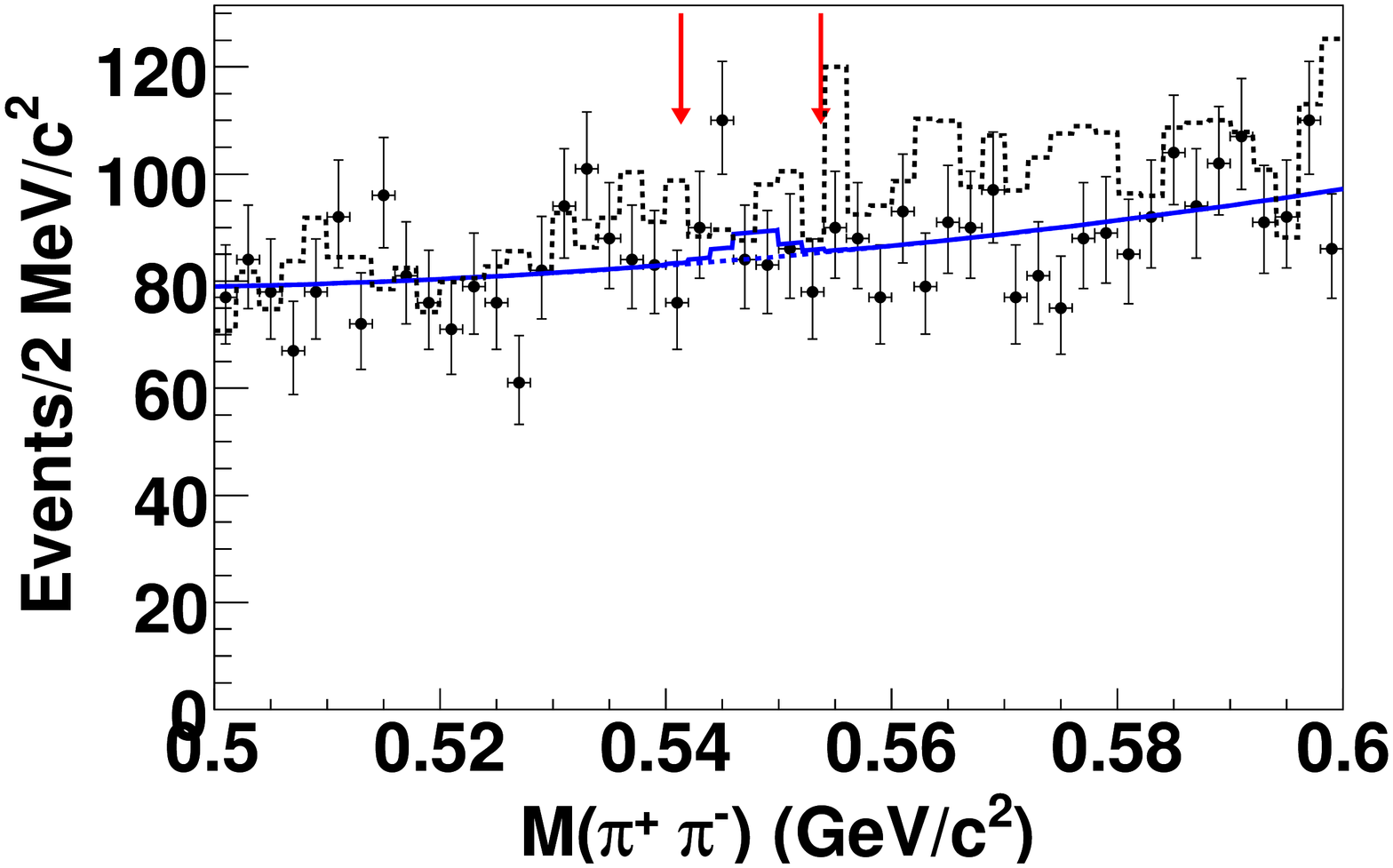}\put(-160,-12){\bf (a)}\\
\includegraphics[angle=-90,width=0.4\textwidth]{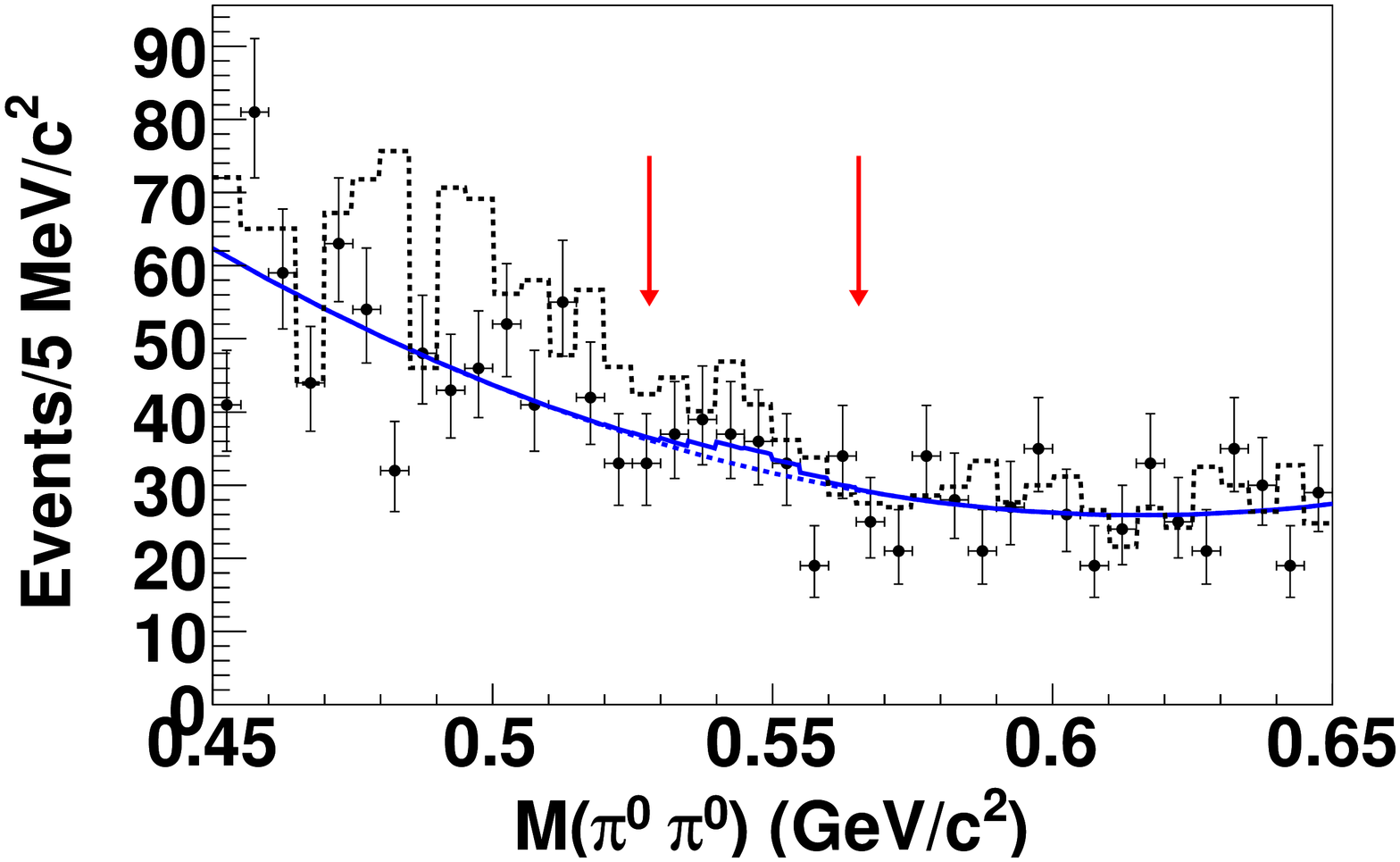}\put(-160,-12){\bf (b)}
\caption{Adapted from Ref.~\cite{BESIII:2011ggk}, which is from $J/\psi\to\gamma\pi\pi$ based
on a sample of $2.2\times 10^{8}$ $J/\psi$ events at BESIII. The $\pi^{+} \pi^{-}$ (a) and
$\pi^{0} \pi^{0}$ (b) invariant mass distributions of the final candidate events in the $\eta$
signal region. The dots with error bars are data, the solid lines are the fit results, and the
dashed histograms are the sum of all the simulated normalized backgrounds. The arrows show mass
regions which contain around 95\% of the signal according to MC simulations.
}
\label{eta_2piBESIII}
\end{figure}

With a sample of $2.2 \times 10^{8}$ $J/\psi$ events, BESIII performed the search
for $\eta\to\pi\pi$ via $J/\psi\to\gamma \eta\to\gamma\pi\pi$ process~\cite{BESIII:2011ggk}.
The dominant background contributions are from $J/\psi\to\pi^+\pi^-\pi^0$, $e^+e^-$ and
$\mu^+\mu^-$ for $\eta\to\pi^+\pi^-$, and $J/\psi\to\gamma\pi^0\pi^0$ with the
direct pions production for $\eta\to\pi^0\pi^0$. In particular the production of
the intermediate state $f_0(600)$ makes the background events irreducible as
illustrated in Fig.~\ref{eta_2piBESIII}. The high background level
makes the sensitivity of searching this rare decay quite low via $J/\psi\to\gamma\eta$,
which set the upper limits as $3.9\times 10^{-4}$ and $6.9\times 10^{-4}$ for
$\eta\rightarrow\pi^+\pi^-$ and $\eta\rightarrow\pi^0\pi^0$, respectively.

\begin{figure}[!htbp]
\centering
\includegraphics[width=0.45\textwidth]{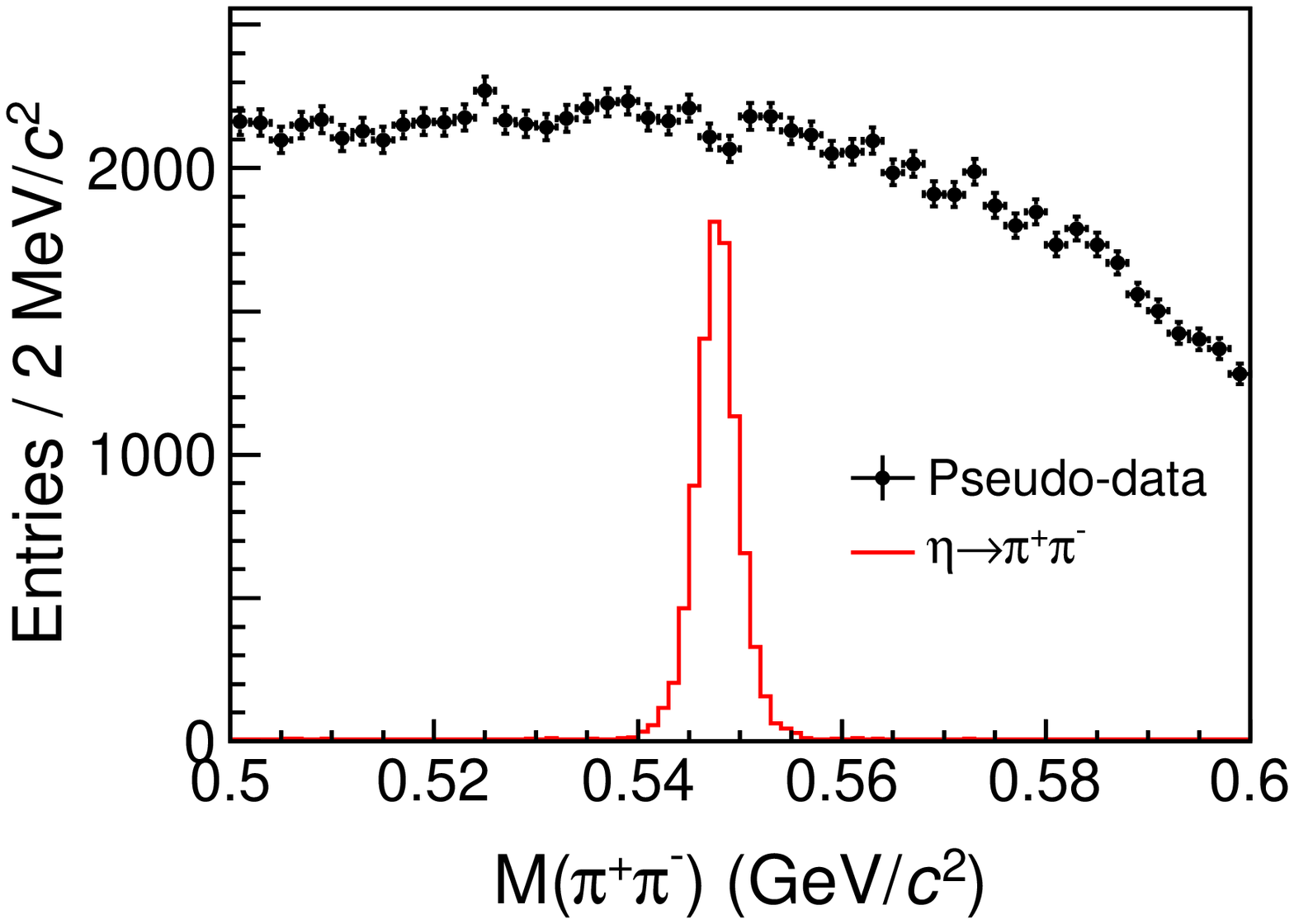}\put(-50,135){\bf (a)}\\
\includegraphics[width=0.45\textwidth]{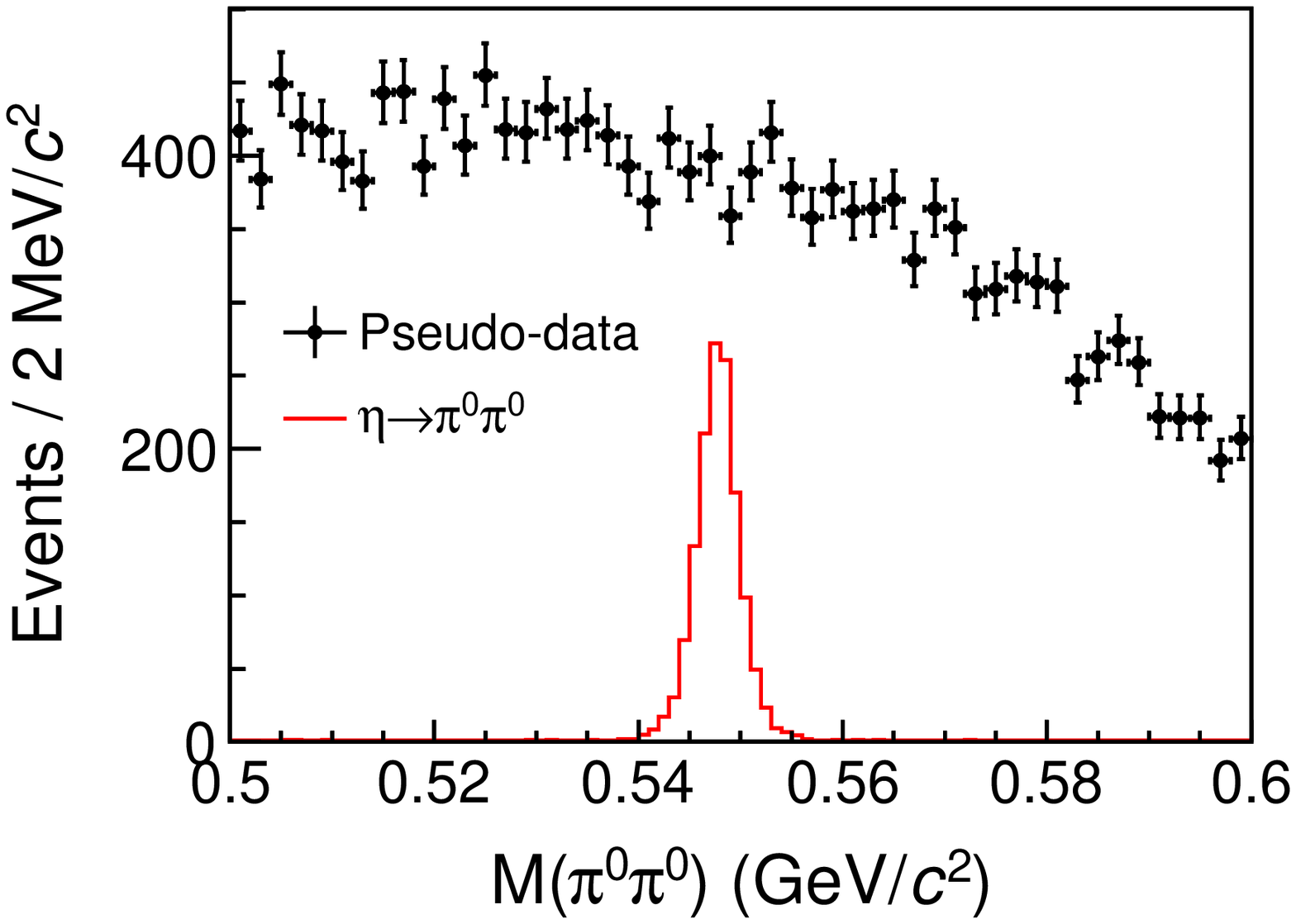}\put(-50,135){\bf (b)}
\caption{The $\pi^{+}\pi^{-}$ (a) and $\pi^{0} \pi^{0}$ (b) invariant mass distributions
for $\eta'\to\pi^+\pi^-\eta(\pi^+\pi^-)$ and $\eta'\to\pi^+\pi^-\eta(\pi^0\pi^0$) candidates
in $\eta$ signal region, respectively. The dots with error bars are from Pseudo-data after
subtracting the non $\eta'\to4\pi$ background contributions and the histograms are the
simulated $\eta\to\pi\pi$ signal with a random scale.
}
\label{eta_2piSTCF}
\end{figure}

To check the sensitivity of searching the rare decay of $\eta\to\pi\pi$ via
$J/\psi\to\gamma\eta'(\pi^+\pi^-\eta)$, MC studies are performed on the Pseudo-data
produced at STCF. The main background events are found to be $\eta'\to\pi^+\pi^-\pi^+\pi^-$
for the charged channel and $\eta^\prime\rightarrow\pi^+\pi^-\pi^0\pi^0$ for the
neutral channel, respectively, which can be well described by the combination of the
ChPT and Vector Meson Dominance (VMD) model. In addition, there are also small amount
backgrounds from $\eta\to\gamma\pi^+\pi^-$ with $\eta$ from $\eta'\to\pi\pi\eta$,
which contribute as peaks in the mass spectra of $\pi^+\pi^-\pi^{+(0)}\pi^{-(0)}$
and also $\pi^+\pi^-$ for the charged channel, but both are below the $\eta'$ and
$\eta$ signal regions. To eliminate $\eta\to\gamma\pi^+\pi^-$ backgrounds and other
continuum background contributions under $\eta^\prime$ peak, the same approach in
Ref.~\cite{BESIII:2014bgm} are adopted. The $M(\pi^+\pi^-)$ or $M(\pi^0\pi^0)$ can
be divided into a number of bins around the $\eta$ signal region and a fit to
$M(\pi^+\pi^-\pi^+\pi^-)$ or $M(\pi^+\pi^-\pi^0\pi^0)$ for each bin is performed
to extract the strength of $\eta^\prime\to4\pi$ and other background contributions.
Then the background-subtracted $\pi^+\pi^-$ and $\pi^0\pi^0$ mass spectra are obtained
and shown in Fig.~\ref{eta_2piSTCF}, together with the possible $\eta\to\pi\pi$ signal
with a random scale. Please note that one $\eta'\to\pi^+\pi^-\pi^+\pi^-$ event contributes
to more than one entry in $M(\pi^+\pi^-)$.

We then made a test by determining the production upper limit of $\eta\rightarrow\pi\pi$ 
using the Bayesian approach. A series of unbinned extended maximum likelihood fits 
is performed to the mass spectrum of $\pi\pi$ with an expected signal. In the fit, 
the line shape of the $\eta$ signal is determined by MC simulation, and the background 
is represented with a second-order Chebychev polynomial. The likelihood distributions 
of the fit are taken as the probability density function directly. The upper limit on 
the number of signal events at the 90\% confidence level (C.L.) corresponds to the
number of events at 90\% of the integral of the probability density function.
Considering the estimated detection efficiency, the upper limits on the branching fraction
of $\eta\to\pi^+\pi^-$ and $\eta\to\pi^0\pi^0$ are determined to be $7.5\times 10^{-8}$
and $6.9\times 10^{-7}$, respectively, which will be the best experimental upper limits
and the one for $\eta\to\pi^0\pi^0$ is three order magnitudes better than the
present upper limit~\cite{pdg2022}.

A full systematic uncertainty evaluation requires both experimental data and full MC
simulation, therefore, we only have a qualitatively discussion below. The possible
systematic uncertainties sources for the upper limits include the number of $J/\psi$,
the intermediate branching fractions, and the event selection. The number of $J/\psi$
can be determined precisely with its hadronic decays, as described in Ref.~\cite{JpsiNum}.
The uncertainties associated with the intermediate process will be taken from PDG.
The uncertainties associated with event selection mainly from the difference between
MC simulation and experimental data in tracking, particle identification, and photon
reconstruction, which can be studied with clean and high statistics control samples
and are still under optimization. The total systematic uncertainty at STCF is expected
to be at the level of several percents or even less, which only has a minor impact on
the sensitivities of $\eta$ rare decays.

\section{$\eta\to\gamma e^+ e^-$ and $\eta\to\gamma\mu^+\mu^-$} 

The $\eta\to\gamma l^+l^-$ ($l=e,~\mu$) decays are the simplest radiative
dilepton decays, also named as the Dalitz decays, where the lepton pair is
formed by internal conversion of an intermediate virtual photon. The deviation
of the spectrum, $M(l^+l^-)$, from the Quantum Electrodynamics (QED) prediction
allows one to investigate the electromagnetic structure of the $\eta$ in terms
of a timelike transition form factor, which has an important role in the evaluation
of the hadronic light-by-light contribution to the muon anomalous magnetic moment.

The latest slope of the form factor measurements for $\eta$ meson are
$\Lambda^{-2}=1.97\pm0.11$ (GeV/$c^2)^{-2}$ and $\Lambda^{-2}=1.934\pm0.067\pm0.050$ (GeV/$c^2)^{-2}$,
respectively, from A2 collaboration using $\eta\rightarrow\gamma e^+e^-$~\cite{Adlarson:2016hpp}
and NA60 collaboration using $\eta\rightarrow\gamma \mu^+\mu^-$~\cite{NA60:2016nad},
while the branching fractions of them have never been updated for more than one decade.

In the study of the $\eta\rightarrow\gamma l^+l^-$ decays with $J/\psi\rightarrow\gamma\eta$
by the BESIII experiment,  it was found that these decays sufferers from the
background events directly from $e^+e^-$ annihilations and $J/\psi$ decays that
have charged pions in the final states. In particular for the $\eta\to\gamma\mu^+\mu^-$
decay, the impact of the backgrounds should be large because of its low branching
fraction and the misidentification of muons and pions. However, the MC study
indicates that both of these two decay modes could be easily distinguished from
events obtained through the $\eta^\prime\rightarrow\pi^+\pi^-\eta$ decay.

\begin{figure}[htbp]
\centering
\includegraphics[width=0.45\textwidth]{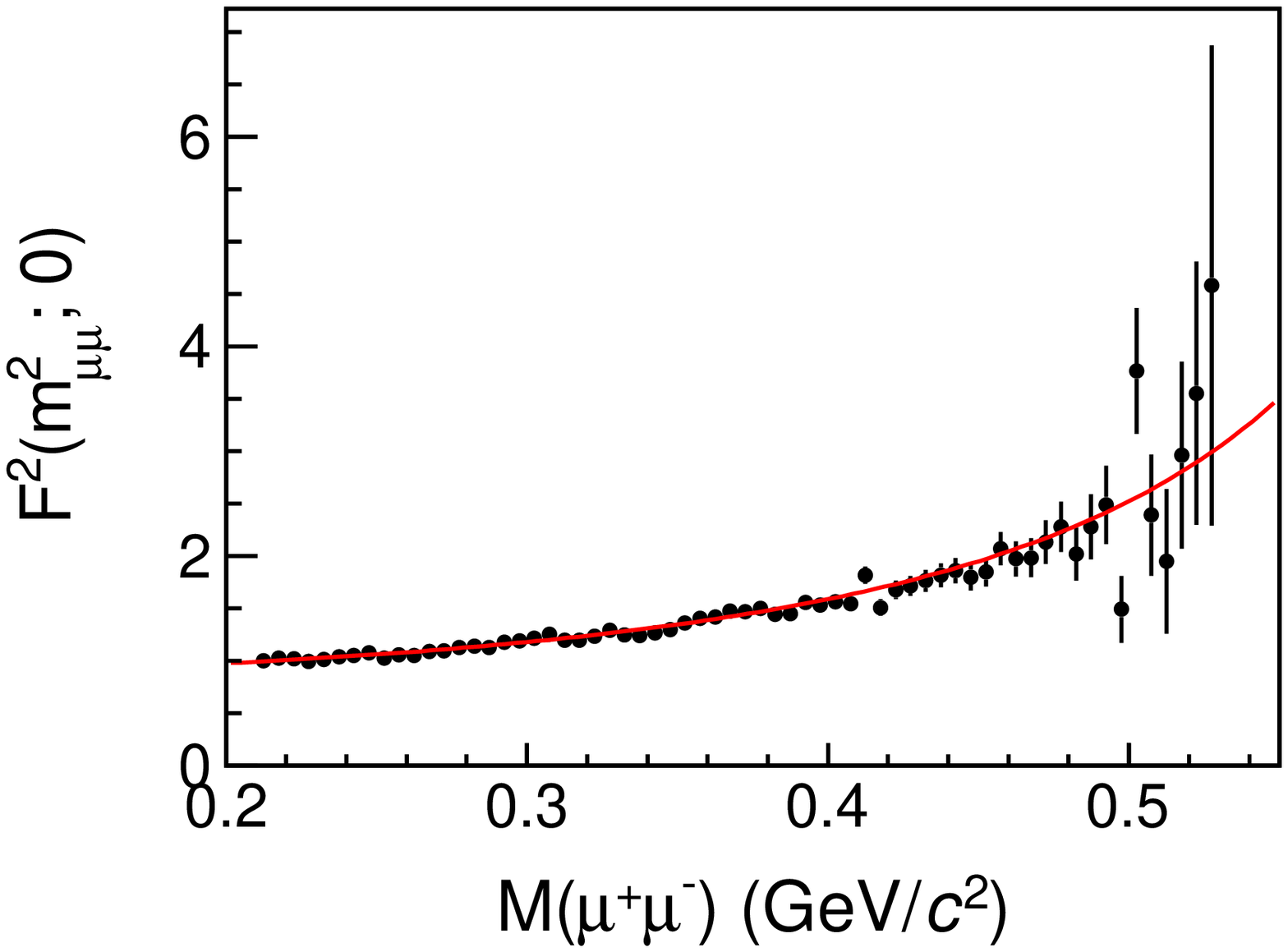}\put(-170,135){\bf (a)}\\
\includegraphics[width=0.45\textwidth]{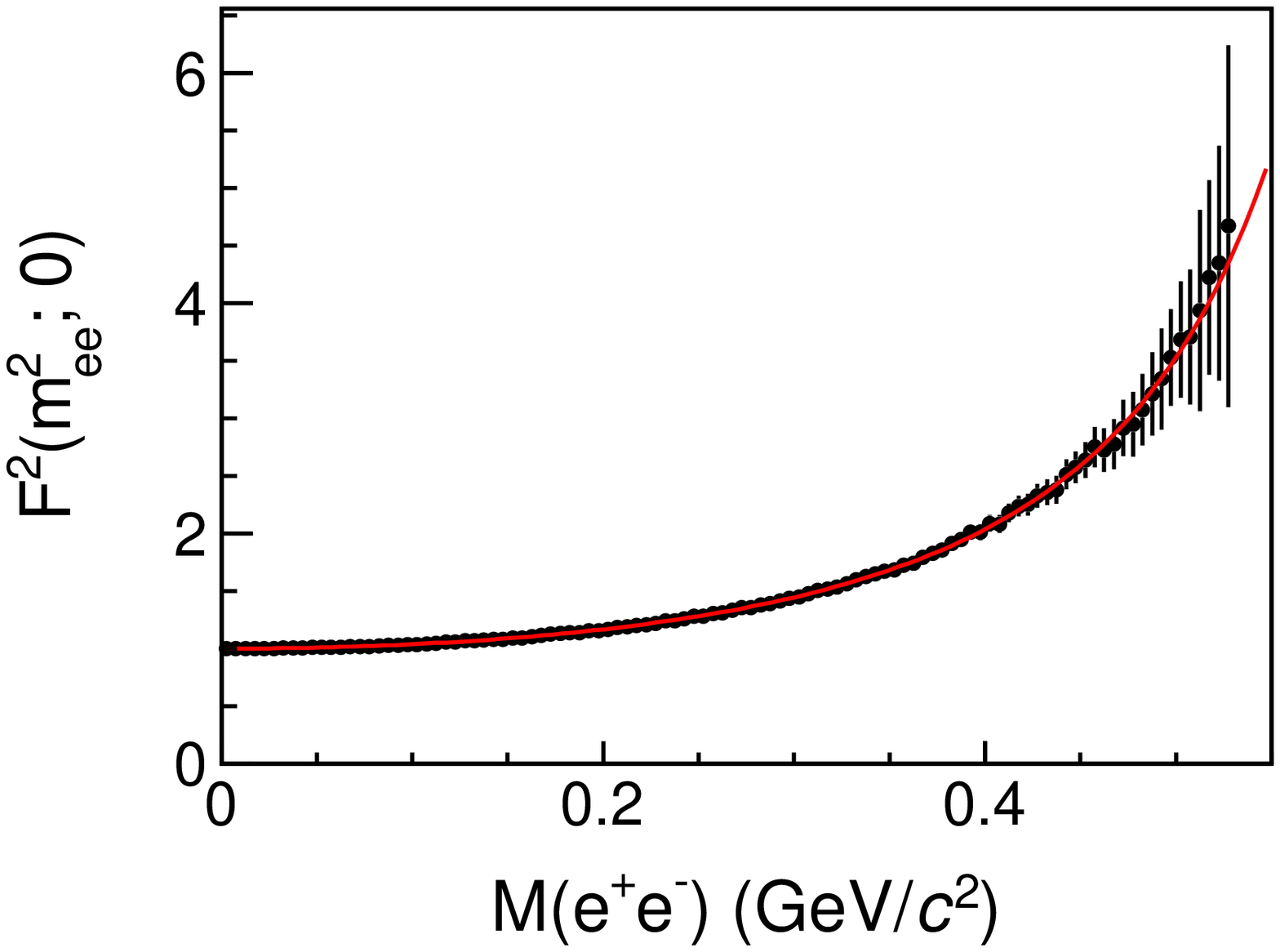}\put(-170,135){\bf (b)}
\caption{The distribution of $F^2(M^2_{l^+l^-};0)$ over the $M(\mu^+\mu^-)$ (a) and
$M(e^+e^-)$ (b). The dots with error bars are the ratio of the background-subtracted
Pseudo-data at STCF to the signal MC which is simulated using $F^2(M^2_{l^+l^-};0)\equiv 1$.
The solid lines are normalized fit results.
}
\label{fig:gammall}
\end{figure}

Using the Pseudo-data at STCF, we selected $200193\pm447$ $\eta\to\gamma e^+e^-$
events and $1747071\pm1321$ $\eta\to\gamma\mu^+\mu^-$ events, respectively.
It was found that the background contribution is at a level of $10^{-3}$, which
indicates that the selected sample of $\eta$ Dalitz decays from
$\eta^\prime\to\pi^+\pi^-\eta$ could provide a clean laboratory to measure the
transition form factor.  After normalization with the QED contribution, the
transition form factors, defined as $F(M^2_{l^+l^-};0)$, as a function of
$M(l^+l^-)$ are displayed in Fig.~\ref{fig:gammall}.  With the single pole model,
$F(M^2_{l^+l^-};0)\equiv (1-M^2_{l^+l^-}/\Lambda^2)^{-1}$, the slopes of the
transition form factor, defined as $dF(M^2_{l^+l^-};0)/dM^2(l^+l^-)=\Lambda^{-2}$,
are measured to be $1.653\pm0.038$ (GeV/$c^2)^{-2}$ for $\eta\to\gamma\mu^+\mu^-$
and $1.644\pm0.012$ (GeV/$c^2)^{-2}$ for $\eta\to\gamma e^+e^-$, where the errors
are statistical only. From the above study, it is clear that the precision of
branching fractions and the transition form factor measurement will be improved
significantly.

In addition, the clean sample of $\eta\rightarrow\gamma\mu^+\mu^-$ allows to
search for the electromagnetic bound states of a $\mu^+\mu^-$ pair, named as
muonium~\cite{bnn969,hb1971}, which, experimentally, has never been observed
yet due to its low production rate.  The observation of the muonium will be
essential for  understanding the various potential anomalies involving muons~\cite{ts2011}
and the possible contributions from the  physics beyond the SM~\cite{cppby2019}.

\section{$\eta\to e^+e^-$ and $\eta\to\mu^+\mu^-$ }

$\eta\to l^+ l^-$ is a fourth order electromagnetic transition and the branching
fraction is expected to be tiny. In particular for $\eta\to e^+e^-$, which is
suppressed compared to $\eta\to\mu^+\mu^-$  as a consequence of the helicity
factor of the electrons. The unitarity limit gives the branching fraction at a
level of  $10^{-9}$~\cite{BABU1982449}, which makes $\eta\to e^+e^-$ an attractive
prospect for a leptoquark search. New theories~\cite{Geng:1990dw,Geng:1990gr} beyond
the SM, such as composite, grand unified and technicolor models, require the existence
of new particles. An especially popular type is known as leptoquark, LQ, which couples
directly to quarks and leptons.   In addition, the interest in the decays was revived
due to the observed excess rate of the $\pi^0\rightarrow e^+e^-$ decay~\cite{a2007}
with respect to the SM predictions~\cite{d2007}. This triggered theoretical speculations
that the excess might be caused by a neutral vector meson responsible for annihilation
of a neutral scalar dark matter particle~\cite{kst2008}. The consequence could be large
(even an oder of magnitude) enhancement of the $\eta\rightarrow e^+e^-$ decay rate.
Therefore, a telling clue to the existence of these new effect would be the enhancement
of $\mathcal{B}$($\eta\to e^+ e^-$) much above the unitary limit, which implies that the
rare decay of $\eta\rightarrow e^+e^-$ can be an important probe for the new physics
beyond the SM.

\begin{figure}[htbp]
\centering
\includegraphics[width=0.45\textwidth]{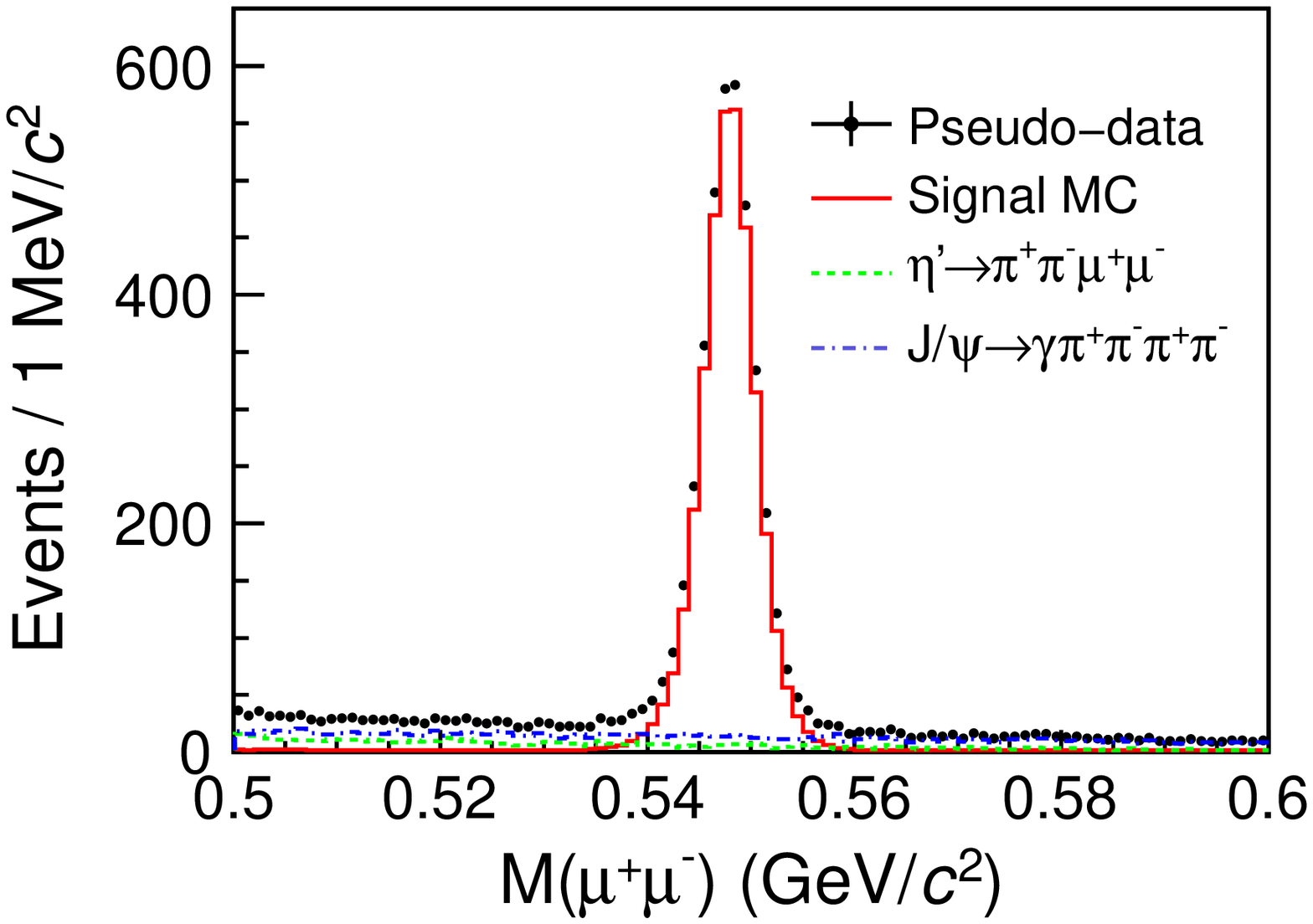}\put(-170,135){\bf (a)}\\
\includegraphics[width=0.45\textwidth]{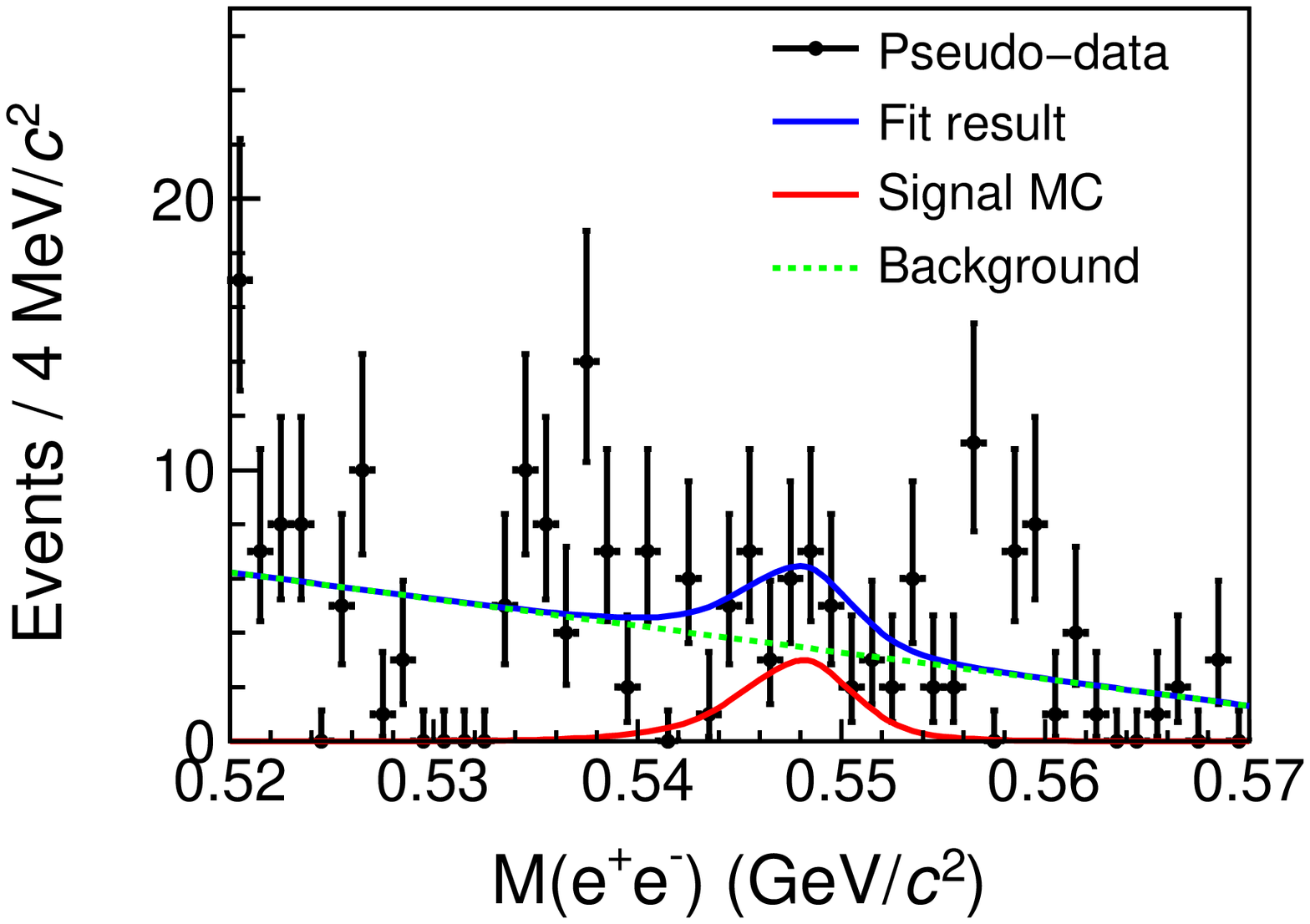}\put(-170,135){\bf (b)}
\caption{The $\mu^{+}\mu^{-}$ (a) and $e^{+}e^{-}$ (b) invariant mass distribution
for $\eta'\to\pi^+\pi^-\eta(\mu^+\mu^-)$ and $\eta'\to\pi^+\pi^-\eta(e^+e^-)$
candidates in $\eta$ signal region, respectively. The dots with error bars are
for the Pseudo-data, the dashed lines are backgrounds, the solid lines are signal
and also fit result for $\eta\to e^+e^-$ channel.
}
\label{etaTomumu_fit2mu}
\end{figure}

Since the high production cross section of $e^+e^-\rightarrow l^+l^-$ and the large
branching fraction of $J/\psi\rightarrow l^+l^-$,  it is hard to investigate
$\eta\rightarrow l^+l^-$ processes using the radiative decay of $J/\psi\rightarrow\gamma\eta$.
However, theoretically the $\eta^\prime\rightarrow\pi^+\pi^- l^+l^-$ decay proceeds
via a virtual photon intermediate state, $\eta\rightarrow \pi^+\pi^-\gamma^*\rightarrow\pi^+\pi^-l^+l^-$.
A peak with a long tail just above 2m$_e$ is expected to be seen in the $M(l^+l^-)$
and a dominant $\rho$ contribution in $M(\pi^+\pi^-)$. These two prominent features
make these decays could be well separated from the decays of
$\eta^\prime\rightarrow\pi^+\pi^-\eta$ with $\eta\rightarrow l^+l^-$, which is illustrated
in Fig.~\ref{etaTomumu_fit2mu}.

Based on 1.3 billion $J/\psi$ events, BESIII first observed $\eta^\prime\to\pi^+\pi^-\mu^+\mu^-$
signal and found a few dozens of events peaked around the $\eta$ meson mass in
the dimuon mass spectrum~\cite{BESIII:2020elh}. These events come from the
$\eta^\prime\to\pi^+\pi^-\eta$, followed by the rare decay $\eta\to \mu^+\mu^-$,
which could give a compatible branching fraction with the present world average value
$\mathcal{B}(\eta\rightarrow\mu^+\mu^-)=(5.8\pm0.8)\times10^{-6}$~\cite{pdg2022}.  
With the current available 10 billion $J/\psi$ events at the BESIII experiments,
which is about eight times larger than that used in Ref.~\cite{BESIII:2020elh},  
the precision of the evaluated branching fraction of $\eta\rightarrow\mu^+\mu^-$
can be extracted with a relative uncertainty of the order of 10\%.

To estimate the background contribution, we performed a MC study by generating
$J/\psi\rightarrow\gamma\eta',\eta'\rightarrow\pi^+\pi^-\mu^+\mu^-$ and
$J/\psi\rightarrow\gamma\pi^+\pi^-\pi^+\pi^-$ samples based on the STCF
fast simulation package, which are also shown in Fig.~\ref{etaTomumu_fit2mu}(b).
Based on the Pseudo-data at STCF, the signal yield of $\eta\rightarrow\mu^+\mu^-$
is estimated to be $3847\pm62$ and the corresponding branching fraction is
calculated to be $(5.88\pm0.09) \times 10^{-6}$, the precision is improved
by one order of magnitude.

With the same Pseudo-data sample, the possible $\eta'\to\pi^+\pi^-\eta$ with
$\eta\to e^+e^-$ candidates are also selected. The obtained $e^+e^-$ mass spectrum
are shown as the blacks dots in Fig.~\ref{etaTomumu_fit2mu}(a). An unbinned maximum
likelihood fit is then performed to the $M(e^+e^-)$ distribution, where the signal
is described by the MC simulated shape, and the background contribution is described
by a first-order Chebychev polynomial function. The branching fraction is expected
to reach a level of $10^{-9}$ with one trillion $J/\psi$ events at STCF, which is
just close to the theoretical calculation. Therefore, an observation of
$\eta\to e^+e^-$ decay with a branching fraction exceeding the theoretical prediction
might be a signature of physics beyond the SM.

\section{$\eta\to\pi^0 e^+e^-$ and $\eta\to\pi^0\mu^+\mu^-$}

\begin{figure}[htph]
\begin{center}
\includegraphics[width=6cm]{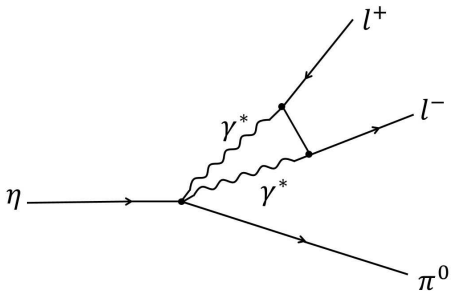}
\caption{$\eta\to\pi^0\gamma^*\gamma^*\to\pi^0 l^+ l^-$ occurring via the $C$-conserving 
second-order electromagnetic process.}
\label{Fig:eepi0}
\end{center}
\end{figure}

The investigation of the charge conjugation invariance in the electromagnetic  
interactions can be done by studying the $\eta\to\pi^0 l^+l^-$ decay. In the
framework of the SM and QED, the matrix element for this process should involve
the two virtual photon exchange~\cite{Smith:1968ab} as it is presented in 
Fig.~\ref{Fig:eepi0} with the transition according to the reaction of 
$\eta \to \pi^0 + \gamma^* + \gamma^* \to \pi^0 + l^+ + l^- $.
The decay rate of those $C$-conserving process, predicted theoretically ranges
from $10^{-11}$ to $10^{-8}$~\cite{Llewellyn:1967tt,Cheng:1967zza,Ng:1993sc}
depending on the undertaken assumptions. Since the first order electromagnetic
$\eta$ decays are forbidden and $\eta\to\pi^0\gamma$ also violates the
conservation of angular momentum, in principle the decay $\eta\to\pi^0 l^+l^-$
proceeds with a virtual photon is forbidden. 

At present, the experimental upper limit for the branching fraction
${\cal B}(\eta\to\pi^0 e^+e^-)$ was determined to be $8\times 10^{-6}$~\cite{pdg2022},
which is still at least three orders of magnitude remains to be experimentally
investigated until reach the prediction based on the SM. While the experimental
upper limit for $\eta\to\pi^0\mu^+\mu^-$, $5\times10^{-6}$~\cite{pdg2022}, has
not been updated for more than 40 years. The observation of any higher branching
fraction than one calculated in the framework of the SM could provide the evidence
that the decay $\eta\to\pi^0 l^+l^-$ is not conserving $C$-invariance.

\begin{figure}[htbp]
\centering
\includegraphics[width=0.45\textwidth]{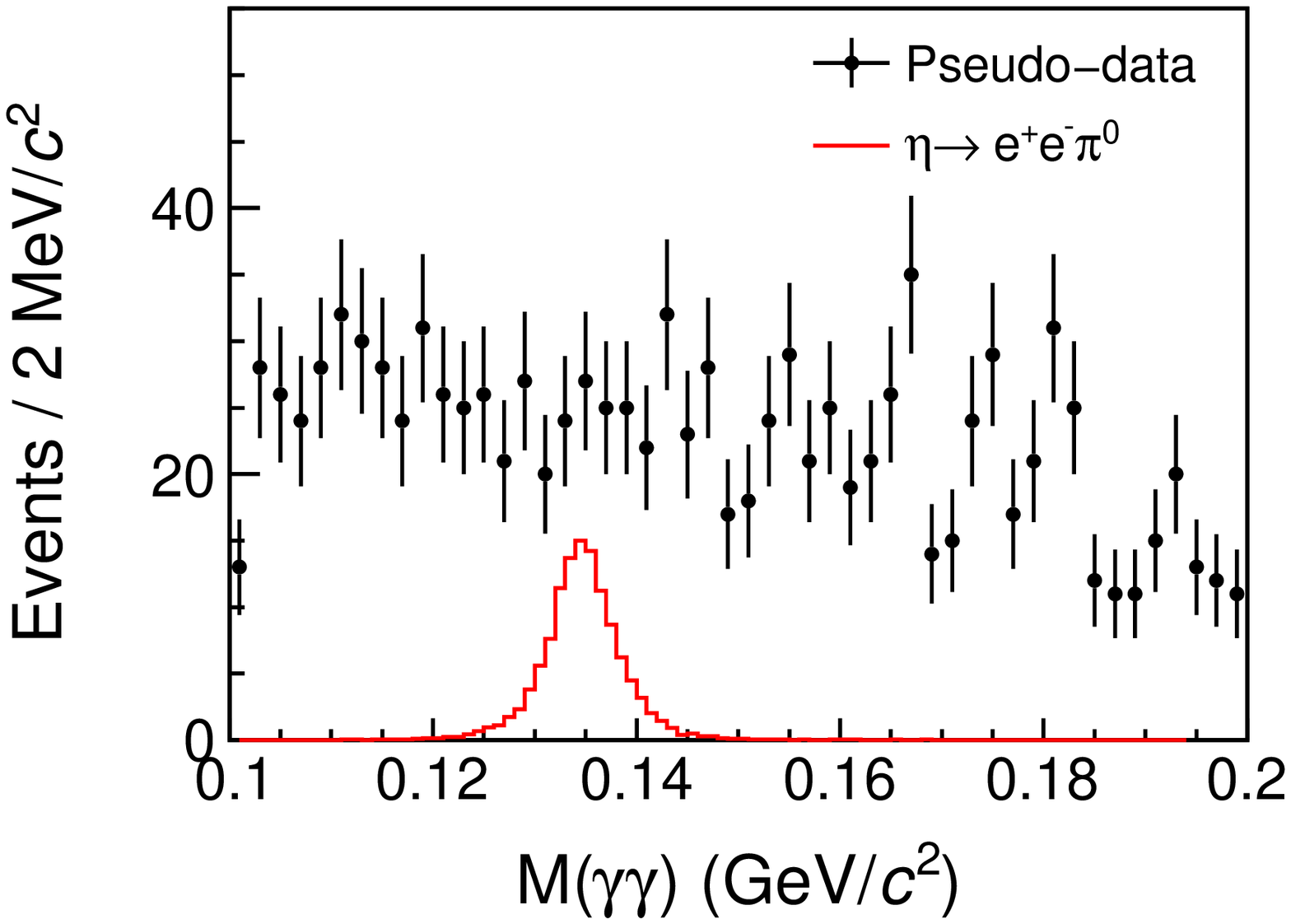}\put(-120,135){\bf (a)}\\
\includegraphics[width=0.45\textwidth]{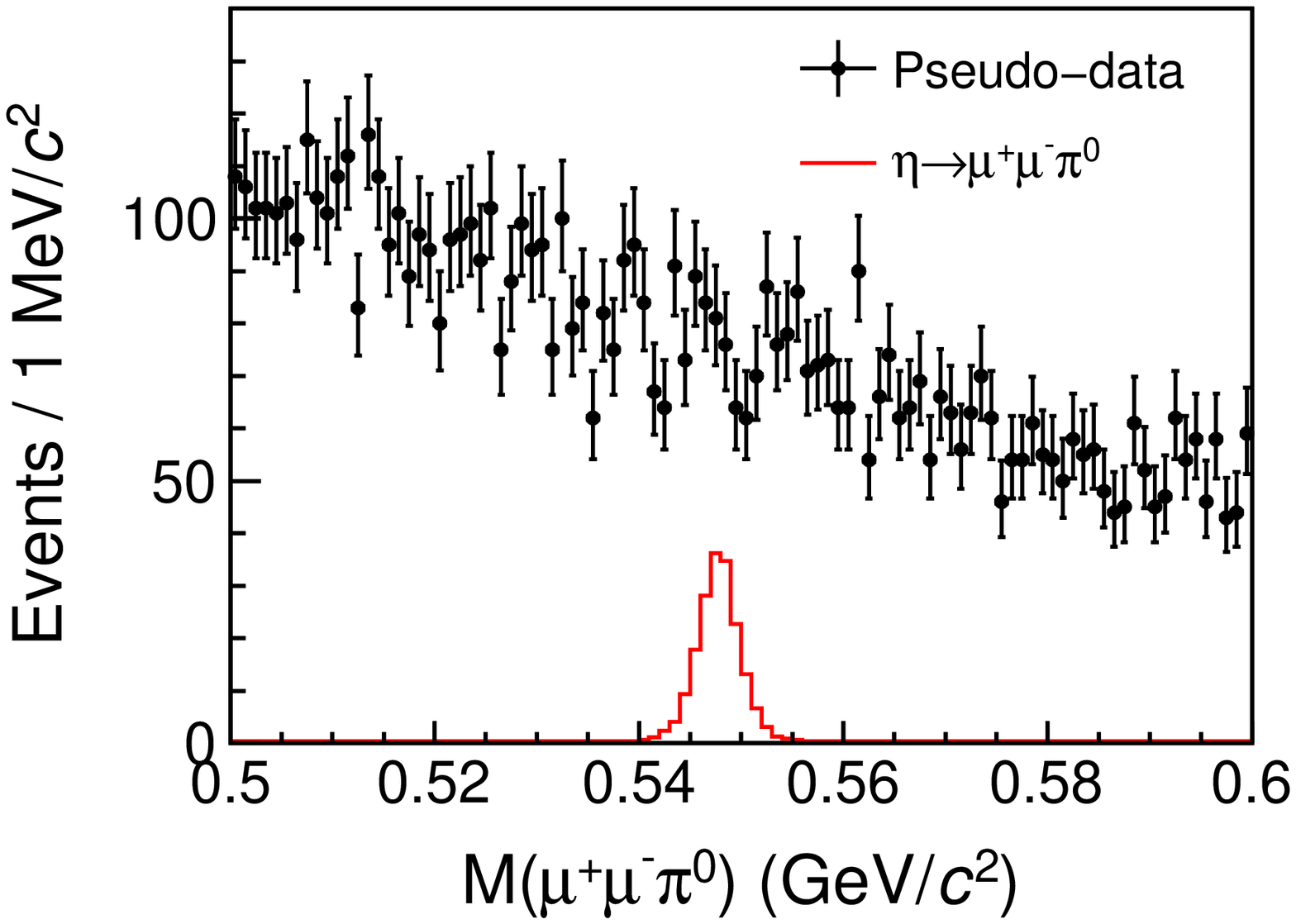}\put(-120,135){\bf (b)}
\caption{(a) The $\gamma\gamma$ mass spectrum with $M(e^{+}e^{-}\gamma\gamma)$
within the $\eta$ signal region. (b) The $\mu^{+}\mu^{-}\pi^{0}$ mass spectrum
for $\eta\to\mu^{+}\mu^{-}\pi^{0}$ channel. The dots with error bars are backgrounds
estimated from the Pseudo-data at STCF and the histograms are the possible
$\eta\to\pi^{0}l^{+}l^{-}$ signal with a random scale.
}
\label{etaTollpi0}
\end{figure}

To testing the feasibility of search for $\eta\to\pi^{0}l^{+}l^{-}$ via
$J/\psi\to\gamma\eta^\prime, \eta^\prime\to\pi^+\pi^-\eta$, studies are
performed with the Pseudo-data sample and the dedicated signal MC samples.
The main backgrounds for the decay process $\eta\to\pi^{0}e^{+}e^{-}$ are
from $\eta\to\gamma e^{+}e^{-}$, which presents as a sharp peak in the mass
spectrum of $\pi^{0}e^{+}e^{-}$ in the $\eta$ signal region, but continuously
in the mass spectrum of $\gamma\gamma$. Therefore, we can easily extract
the possible $\eta\to\pi^{0}e^{+}e^{-}$ signal by fitting to the mass spectrum
of $\gamma\gamma$ with the requirement of $M(e^{+}e^{-}\gamma\gamma)$ in $\eta$
signal region. Fig.~\ref{etaTollpi0}(a) shows the obtained $\gamma\gamma$ mass
spectrum from the Pseudo-data sample and the possible $\eta\to\pi^{0}e^{+}e^{-}$
signal with a random scale. With one trillion $J/\psi$ events at STCF, the upper
limit is expected around $2\times10^{-7}$, which is improved by one order magnitude
compared with the PDG value~\cite{pdg2022}.

While for $\eta\to\pi^{0}\mu^{+}\mu^{-}$ channel, the main backgrounds are
from $\eta\to\pi^{+}\pi^{-}\pi^{0}$, which is flat in the mass spectrum of
$\mu^{+}\mu^{-}\pi^{0}$ around $\eta$ signal region. Fig.~\ref{etaTollpi0}(b)
shows the background contributions estimated from the Pseudo-data sample
and the possible $\eta\to\pi^{0}\mu^{+}\mu^{-}$ signal with a random scale.
By fitting to $M(\mu^{+}\mu^{-}\pi^{0})$, we can estimate the possible
$\eta\to\pi^{0}\mu^{+}\mu^{-}$ signal yields. Together with the estimated
efficiency, the upper limit on the branching fraction for $\eta\to\pi^{0}\mu^{+}\mu^{-}$
is expected to reach $8.5\times10^{-8}$ with one trillion $J/\psi$ events at STCF,
which is improved by two order magnitudes compared with the PDG value~\cite{pdg2022}
and quite close to the theoretical prediction.

\section{Summary}

Despite the impressive progresses on the investigation of $\eta$ mesons
were achieved recent years, the data on the decay modes of the  $\eta$
are still scarcer and much less accurate than those for the pions and kaons. 
The reason is that the $\eta$ mesons were produced with low intensity,  
which inspired new facilities proposed for dedicating to explore the
$\eta/\eta^\prime$ decays~\cite{JFeta, REDTOP}. Moreover, the STCF is
unique since the charmonium decays ($J/\psi$) provides a very clean light
meson samples as advocated by the BESIII experiment~\cite{stcf}.

For the investigation on the $\eta$ decays,  since its production rate
of  $J/\psi\rightarrow\gamma \eta$  is five times less than that of
$\eta^\prime$ in $J/\psi$ radiative decays and the irreducible background
contributions directly from both $J/\psi$ decays and $e^+e^-$ annihilations,
it is hard to improve the sensitivity for exploring the $\eta$ rare decays.
However,  $\eta^\prime\rightarrow\pi^+\pi^-\eta$ is one of dominant decays
with a branching fraction of $(42.5\pm0.5)$\%~\cite{pdg2022} and the $\eta$
mesons could be well tagged, these features make the decay of
$\eta^\prime\rightarrow\pi\pi\eta$ particularly attractive for the study
of $\eta$ decays, which inspired us to present a proposal for exploring the
$\eta$ decays by tagging $\eta$ with $\eta^\prime\rightarrow\pi^+\pi^-\eta$
at the STCF~\cite{stcf}.

STCF was proposed to perform an extensive study of $\tau$-charm physics~\cite{stcf}
and the designed luminosity is about 100 times larger than that of BEPCII.
Therefore, the unprecedented charmonium decays, e.g., $J/\psi$ and $\psi(2S)$,
are expected to be accumulated in one year. We then present several examples
of physics feasibility studies performed with the fast simulation package
developed for STCF. The examples are not intended to deliver an applicable
message for this novel approach, instead, they are provided to illustrate
the STCF capabilities to fulfill this physics program. The MC study indicates
that STCF  opens the possibility to investigate the $\eta$ decays with an
excellent sensitivity and may make feasible observation of $\eta$ rare decays.
Actually, the above study also advocates that the available 10 billion $J/\psi$
events~\cite{BESIII:2021cxx} can already yield a series of  measurements, such
as $\eta\rightarrow 2\pi$ and $\eta\rightarrow l^+l^-\pi^0$, with accuracy
competitive with the current world averages.

\begin{acknowledgments}
We thank the Hefei Comprehensive National Science Center for their strong support
on the STCF key technology research project. This work is supported by 
National Natural Science Foundation of China (NSFC) under Contracts No. 12005195, No. 12225509,
the National Key R\&D Program of China under Contract No. 2022YFA1602200, the
international partnership program of the Chinese Academy of Sciences Grant No. 211134KYSB20200057,
and Wuhan Scientific Research Project under Contract No. 20231250048.
\end{acknowledgments}

\bibliographystyle{unsrt}

\end{document}